\documentstyle[aps,prl,twocolumn]{revtex}

\begin{document}
\draft

\title{The Metallic-Like Conductivity of a Two-Dimensional Hole System}

\author{Y. Hanein$^{1,2}$, U. Meirav$^{1}$, D. Shahar$^{1,2}$, C.C. Li 
$^{2}$, D.C. Tsui$^{2}$ and Hadas Shtrikman$^{1}$}

\address{$^{1}$ Dept. of Condensed Matter Physics, Weizmann Institute, 
Rehovot 76100, Israel}

\address{$^{2}$ Dept. of Electrical Engineering, Princeton University, 
Princeton, New Jersey 08544}

\maketitle

\begin{abstract}

We report on a zero magnetic field transport study of a two-dimensional, 
variable-density, hole system in GaAs. As the density is varied we 
observe, for the first time in GaAs-based materials, 
a crossover from an insulating behavior at low-density, to a 
metallic-like
behavior at high-density, where the metallic behavior is 
characterized by a large drop in the resistivity as the 
temperature is lowered. These results are in agreement with recent 
experiments on Si-based two-dimensional systems by Kravchenko {\em et al}. \cite{SVKrav94} 
and others \cite{DPopovic97,Ismail,Colridge,JLam97}. We show that,
in the metallic region, the resistivity is dominated by an 
exponential temperature-dependence with a characteristic temperature 
which is proportional to the hole density, and appear to reach a constant 
value at lower temperatures.
\end{abstract}


\date{\today}
The study of the transport properties of two-dimensional electron systems (2DES's) 
revealed numerous unique features
associated with their reduced dimensionality. 
A central question is whether a metal-insulator transition (MIT) can occur in two-dimensions
(2D). 
Using scaling arguments Abrahams et
al.\cite{Gang4} stated that non-interacting electrons in 2D systems  are 
localized at zero temperature ($T$)
for any level of disorder, and a MIT is not expected to occur at zero 
magnetic-field ($B$).
This work motivated several experimental studies which investigated the 
$T$ dependence of the resistivity ($\rho$) of low-mobility 
2DES in Si metal-oxide semiconductor filed-effect transistors (MOSFET's)
\cite{DJBishop80,MJUren81} and in In$_{2}$O$_{3-x}$ films \cite{ZOvadyahu85}.  
The resistivity was found to increase with
decreasing $T$, and its $T$-dependence changed from weak
to strong as the density of the 2DES was lowered, or the disorder 
increased. These experiments 
seemed to confirm the theoretical notion that no metallic phase exists 
in 2D. 

However, several recent
studies presented evidence to the contrary.
In these studies 2DES's in Si MOSFET's \cite{SVKrav94,DPopovic97} and Si/SiGe heterostructure 
\cite{Ismail,Colridge,JLam97} were used, and the resistivity was measured as
a function of $T$ for various carrier-densities.
These studies demonstrated a clear crossover from metallic
to insulating behavior at low $T$. 
Further, in the metallic region, the resistivity was
shown to decrease with decreasing $T$ by as much as a factor of eight, 
while in the insulating region the
resistivity increases sharply with decreasing $T$. 
The authors of refs. \cite{SVKrav94} took these results as 
evidence for the existence of a MIT in 2D, and several theoretical 
works have tried to associate them with modified scaling \cite {VDobrosavljevic97} 
or raised the possibility of superconductivity \cite{PPhillips97,DBelitz97}.
The disagreement between these new results and earlier ones are 
generally attributed to the higher mobility in these 
samples (reaching a value as high as $75,000$ cm$^2/V{\cdot}s$) and to the high 
effective mass of electrons in Si ($m=0.19m_0$) which, according 
to the authors, combine to 
accentuate the effect of carrier-carrier interactions. 
It was therefore suggested \cite{SVKrav97} that  
due to the heavy mass of holes in GaAs ($0.6m_{0}, 0.38m_{0})$ 
\cite{HLStormer83,KHirakawa93} and the superior quality of molecular beam epitaxy (MBE) growth, 
a similar
transition may appear in two-dimensional hole system (2DHS) embedded in 
GaAs heterostructures. 

In this letter we report on a low-$T$, $B=0$, 
study of high-quality 2DHS in GaAs that indeed exhibits such a 
transition.  
To facilitate this study we chose a 
p-type inverted semiconductor insulator semiconductor (ISIS) 
structure \cite{UMeirav88} 
grown on (311)A GaAs substrate \cite{YHanein97} using Si as a $p$-type dopant \cite{LPavesi95}. 
A schematic layer-profile of the 
$p$-type ISIS structure used in this work is shown in the inset of Fig 1. 
In such ISIS device the carriers are
accumulated in an undoped GaAs layer lying on top of an undoped AlAs barrier,
grown over a $p^+$ conducting layer. This $p^{+}$ conducting layer 
is separately contacted and serves as a back-gate.
The hole carrier-density ($p$) can
be easily varied by applying voltage ($V_{g}$) to the back-gate resulting 
in a change in $p$ of $1.1{\cdot} 10^{11}$ cm$^{-2}V^{-1}$, 
which is consistent with a capacitively induced
charge-transfer. The carrier density was determined by direct 
measurement of the Hall effect. At $T=50$ mK our samples have a 
mobility of 150,000 cm$^{-2}/V{\cdot}s$ at $p=0.64{\cdot}10^{11}$ 
cm$^{-2}$. The high limit of the density in this study was set by gate leakage.
The samples were wet-etched to the shape of a
standard Hall-bar and Zn-Au evaporated contacts were carefully alloyed at $370^{\circ}$C 
to avoid penetration of the alloyed metal into the $p^+$ buffer layer. 

One handy advantage of the ISIS structure is that it offers the capability to vary the carrier density
continuously over a very wide range, in a similar fashion to Si MOSFET's. We utilize this
flexibility to study the $T$-dependence of $\rho$ at a number of densities from the insulating
phase to the high-$p$ conducting state. 
Measurements were done in a dilution refrigerator with a base $T$ of $40$
mK, using AC lock-in technique with an excitation current of $0.1$ nA.
We repeated the measurements with higher excitation current and no effect of heating 
was found.
The collection of the data was done by fixing the density and sweeping
$T$ continuously between our base $T$ and 1.2 K.  The data were reproducible 
upon numerous cycles
of $V_g$ and $T$.

Typical results, obtained from sample H315J, are
presented in Fig. 1, where we plot $\rho$ vs. $T$ at various values 
of $p$ between $0.089{\cdot}
10^{11}$ and $0.64{\cdot}10^{11}$ cm$^{-2}$. It is possible to crudely classify the 
traces in Fig. 1 into three distinct regimes. 
The first is the low-$p$ regime (top set of solid lines), characterized by insulating
behavior, with $\rho(T)$ decreasing rapidly with $T$. In the second regime (dashed curves), a mixed
behavior is observed at our $T$-range, with insulating-like $\rho(T)$ at 
high-$T$, turning over
to metallic-like behavior at lower $T$'s. 
Although quite interesting, we defer discussion of these
two regimes to a future publication, and center our attention on the 
high-$p$ region,
where the $\rho(T)$ traces exhibit metallic-like behavior characterized by $\rho$ 
that drops precipitously as $T$ is lowered, followed by an apparent saturation of $\rho(T)$ at
yet lower $T$'s. 

The results presented in Fig. 1 are very similar to those
of Kravchenko {\em et al}. \cite{SVKrav94} and others \cite{DPopovic97}, 
a notable fact considering that we are dealing with two
distinct materials. Incidentally, this similarity rules out explanations of the
data that are material dependent.
We wish to point out that in both Si MOSFET's and p-type GaAs ISIS the low carrier-density,
coupled with their high effective mass, lead to a relatively low Fermi 
energy ($E_F$) and high ratio of the typical carrier-carrier Coulomb-interaction energy to 
$E_{F}$, expressed by the dimensionless parameter $r_{s}$. In our 
density range, $E_F=0.5$ to 3.7 K and $r_{s}=24$ to 9 calculated for the lighter 
mass, $m=0.38m_{0}$. 
It is not clear whether we can safely deduce the
ultimate low-$T$ phase of our system from measurements done at $T$'s that are not much smaller
than $E_F/k_B$.

Of their many findings, Kravchenko {\em et al}. \cite{SVKrav94} 
singled out the existence of an
apparent  metallic behavior in their samples as the most intriguing. We therefore turn our
attention to a detailed analysis of the high-$p$ range of our data where 
metallic-like behavior is evident. 
Following the suggestion of Pudalov \cite{VMPudalov97}, we fit the $\rho(T)$ data with
\begin{equation}
\rho(T)=\rho_0+\rho_1\exp(-\frac{T_0}{T})
\label{PudFit}
\end{equation}
and plot, in Fig. 2, $\rho$  vs. $T$ (solid lines) 
for several values of $p$ in the metallic regime, 
together with their fits (dashed lines). A surprisingly good fit (at 
low $T$) is obtained especially at high-$p$, where only marginal and 
hardly noticeable deviations of the fits from the data are seen.
In the remainder of this
letter we will explore this form and its possible consequences.

We begin by considering the limits of applicability of Eq. \ref{PudFit}. 
At higher $T$'s we observe significant deviations from the behavior prescribed 
by Eq. \ref{PudFit} and indicate 
their approximate $T$'s by arrows in Fig. 2. These deviations may be 
expected when $T>T_{0}$. Several other factors may also combine to 
cause additional deviations, among them the aforementioned approach 
towards $E_{F}$ and the influence of the insulating regime at low-$p$. 

More relevant to the determination of the quantum nature of the 
transport is the low-$T$ range. Following Eq. \ref{PudFit}, $\rho$ 
seems to approach a constant, $\rho_0$. We note that this apparent 
saturation is not likely to be caused by heating since, at the same 
$T$ range, lower-$p$ traces (Fig. 1), where the dissipation induced by 
the constant probing-current is more than 10 times larger, are distinctly $T$ dependent.
At face value, a finite $\rho_0$ may 
indicate a true metallic zero $T$ behavior. We stress that this is not 
necessarily the case for finite-size samples, a fact that 
had been previously recognized regarding high-mobility 2DES's. 
Furthermore, due to the limited accuracy of our measurements, 
we are unable to exclude the existence of additional terms in 
$\rho(T)$ with relatively weak (logarithmic) $T$-dependence. 
Moreover, the existence of the exponential term in $\rho$ will 
effectively mask any possible $T$-dependence of $\rho$ as long as $T$ 
is not much smaller than $T_{0}$. We are therefore in a situation 
where our ability to extrapolate $\rho(T)$ to $T=0$ is limited to a 
very narrow $T$ range between $\approx T_{0}/5$ (where the exponential 
term becomes negligible) and our base $T$, 40 mK.  Clearly, 
much lower $T$'s are essential for unveiling the ultimate zero $T$ behavior of the system. 

Between these two extremes of Eq. \ref{PudFit}, a remarkable behavior is observed. 
In this intermediate $T$ range, the
second term of Eq. \ref{PudFit} dominates and $\rho$ is exponentially increasing with $T$. To
further emphasize this behavior we plot, in Fig. 3, $\rho-\rho_0$ vs. $1/T$ using an Arrhenius
plot, for the data of Fig. 2. A clear exponential dependence is seen which 
covers over 2 orders of magnitude in $\rho$. It is this exponential behavior which underlies the abrupt drop
in $\rho$ seen in Figs. 1 and 2. 

Clearly, the physical origin of $T_0$ is one of the main issues that need to be
addressed. We therefore plot, in Fig. 4, the characteristic $T$ of the 
conduction process, $T_0$, as a function of
$p$ for our sample. $T_0$ increases monotonically with $p$ and reaches $0.78$ K for our highest
$p$. With the limited accuracy of these measurements we are unable to determine the functional
dependence of
$T_0$ on $p$, although a linear dependence is reasonable.
We note that from the linear fit  $T_0$ extrapolates to  $T_0$=0 as $p$ 
approaches zero. This implies that the exponential $T$ dependence 
of $\rho$ persists to $p=0$. In reality it is overcome by the 
insulating behavior that sets-in at low-$p$.

At present, we do not have a theoretical understanding of the physics which underlies this
intriguing regime.  
We note that this behavior reveals 
itself under the condition of low $E_{F}$, where
carrier-carrier interactions should play a more significant role in the 
dynamics of the system. 
This may explain why in systems with lower $r_{s}$, such as 2DEG in GaAs, no analogous 
behavior is observed \cite{AJDahm}.  
The observation of an exponential term in $\rho$ may be taken as 
evidence for
the existence of an energy gap, which inhibits scattering at low $T$. 
Pudalov suggested a
spin-orbit gap as the origin of $T_0$ but, as he stated, this gap is 
likely to be too small in GaAs \cite{VMPudalov97}.

To summarize, 
we used a p-type ISIS to investigate the $T$-dependence of the resistivity 
of two-dimensional hole gas in GaAs.
We observed a transition from metallic to insulating behavior 
at low $T$'s. 
The results in this letter are similar to other 
experiments facilitating different 2D systems. 
We have shown that the resistivity at the metallic phase is determined by two terms. 
The first appears to be independent on $T$, while the second factor has a rather 
surprising, exponential, dependence on $T_{0}/T$ and $T_{0}$ increases monotonically with $p$. 
We emphasize that much lower $T$'s are needed to safely determine the 
ultimate low-$T$ phase of the system.

The authors wish to acknowledge very useful discussions with O. Agam and A. Stern.
This work was supported by a grant from the Israeli Ministry of Science and The
Arts and by the NSF.

\begin{figure}[tbp]
\caption{$\rho$ as a function of $T$ data obtained
at $B=0$ at various fixed densities,  
$p=$0.089, 0.94, 0.99, 0.109, 0.119, 0.125, 0.13, 0.15, 0.17, 0.19 
0.25, 0.32, 0.38, 0.45, 0.51, 0.57 and 0.64$\cdot 10^{11}$ cm$^{-2}$. 
Note the three distinct regimes: insulating regime at low densities, 
a mixed regime at intermediate densities indicated by dashed-lines, and
a metallic-like regime at high-densities (see text). 
Inset: Schematic presentation of the 
$p$-type ISIS structure grown on semi-insulating (311)A GaAs
substrate, consisting of a thick $p^+$ buffer, a 300 nm undoped AlAs barrier, 
a 150 nm undoped GaAs
channel layer,  and a top 50 nm GaAs layer which is $p$ doped. The 2DHS forms at the lower
interface of the channel  layer upon application of negative bias to 
the $p^{+}$ conducting layer.}
\label{KLZ}
\end{figure}
\begin{figure}[tbp]
\caption{Metallic-like curves at 
$p=0.30$, 0.36, 0.43, 0.50, 0.57 and $0.64\cdot 10^{11}$ cm$^{-2}$. plotted using a linear 
ordinate (solid lines). 
Also shown are the fits to the empiric formula:
$\rho_{1}+\rho_{1}\exp \left( -\frac{T_{0}}{T}\right)$ (dashed lines).
Due to the high fidelity of the fits, they are hardly distinguishable 
from the data at some higher densities.  Arrows indicate the 
approximate $T$ where the fits start to deviate significantly from the 
data.}
\label{KLZ1}
\end{figure}

\begin{figure}[tbp]
\caption{$\rho$-$\rho_{0}$ ($\rho_{0}$ is obtained from fitting the data to Eq.
\ref{PudFit}) as a function of $T^{-1}$ at same $p$ values as in Fig. 
2. 
The $T$ was varied between 50 mK and 1.2 K.}
\label{KLZ4}
\end{figure}

\begin{figure}[tbp]
\caption{The fitting-parameter $T_{0}$ as a function of $p$.  
The $T_{0}$'s are obtained by fitting the data to Eq. \ref{PudFit}.
The solid line indicates the linear fit of $T_{0}$. 
The dashed line indicates the extrapolation 
of the fit to $p$ equals zero.}
\label{KLZ2}
\end{figure}


\begin{references}

\bibitem{SVKrav94}  S. V. Kravchenko, G. V. Kravchenko, J. E. Furneaux, V. M. Pudalov, and M. D'Iorio,
Phys. Rev. B {\bf 50}, 8039 (1994);
S. V. Kravchenko, W. Mason, G. E. Bowker, J. E. Furneaux, V. M. Pudalov, 
and M. D'Iorio, 
Phys. Rev. B {\bf 51}, 7038 (1995);
S. V. Kravchenko, D. Simonian, M. P. Sarachik, W. Mason, and J. E. 
Furneaux, Phys. Rev. Lett. {\bf 77}, 4938 (1996).

\bibitem{DPopovic97} 
D. Popovic, A. B. Fowler, and S. Washburn,  
Phys. Rev. Lett {\bf 79}, 1543 (1997).

\bibitem{Ismail} 
K. Ismail, J.O. Chu, Dragana Popovic, A.B. Fowler, and S. Washburn,  
preprint cond-mat/9707061 (1997).

\bibitem{Colridge} 
P.T. Coleridge, R.L. Williams, Y. Feng, and P. Zawadzki,  
preprint cond-mat/9708118 (1997).

\bibitem{JLam97} 
J. Lam, M. D'Iorio, D. Brown and H. Lafontaine,  
preprint cond-mat/9708201 (1997).

\bibitem{Gang4}  E. Abrahams, P. W. Anderson,  D. C. Licciardello, and T. Ramakrishnan, Phys.
Rev. Lett. {\bf 42}, 673 (1979).

\bibitem{DJBishop80}  D. J. Bishop, D. C. Tsui,  and R. C. Dynes, Phys.
Rev. Lett. {\bf 44}, 5737 (1980).

\bibitem{MJUren81}  M. J Uren, R. A. Davies,  M. Kaveh, and M. Pepper, J. Phys.
C. {\bf 14}, 5737 (1981).

\bibitem{ZOvadyahu85}  Z. Ovadyahu, and Y. Imry, J. Phys. {\bf C16}, 
L471 (1983).

\bibitem{VDobrosavljevic97} 
V. Dobrosavljevic, Elihu Abrahams, E. Miranda, and Sudip Chakravarty,
Phys. Rev. Lett. {\bf 79}, 455 (1997).

\bibitem{PPhillips97}
P. Phillips and Y. Wan, preprint cond-mat/9704200 (1997).

\bibitem{DBelitz97}
 D. Belitz, T. R. Kirkpatrick, cond-mat/9705023 (1997).
 
\bibitem{SVKrav97}  S. V. Kravchenko (private communication).

\bibitem{HLStormer83}  H. L. Stormer, Z. Schlesinger, A. Chang, D. C. Tsui,  
A. C. Gossard, and W. Wigmann, Phys.
Rev. Lett. {\bf 51}, 126 (1983).

\bibitem{KHirakawa93}  K. Hirakawa, Y. Zhao, M. B. Santos, M. Shayegan, 
and D. C. Tsui, Phys. Rev. B. {\bf 47}, 4076 (1993).

\bibitem{UMeirav88}  U. Meirav, M. Heiblum, and F. Stern, Appl. Phys.
Lett. {\bf 52}, 1268 (1988).

\bibitem{YHanein97}  Y. Hanein, H. Shtrikman, and U. Meirav, Appl. Phys.
Lett. {\bf 70}, 1426 (1997).

\bibitem{LPavesi95}  L. Pavesi, M. Henini, and D. Johnston, Appl. Phys.
Lett. {\bf 66}, 2846 (1985).

\bibitem{VMPudalov97}  V. M. Pudalov, preprint cond-mat/9707076 (1997).

\bibitem{AJDahm}  A. J. Dahm, F.W. Van Keuls, H. Mathur, and H. W. 
Jiang, Proc. Intern. Conf. on Electron Localization and Quantum 
Transport in Solids, Jaszowiec, Poland, 1996. (Ed. by T.Dietel), Inst. 
of Physics PAN, Warsaw (1996).

\end{references}
\end{document}